\documentclass{article}
\usepackage{amssymb,amsmath,latexsym,amscd,amsfonts}
\usepackage{graphics}
\usepackage{epsfig}

\newtheorem{thm}{Theorem}[section]

\newtheorem{definition}{Definition}[section]

\newtheorem{lem}{Lemma}[section]

\newtheorem{corol}{Corollary}[section]

\def\NP{{\sf{NP}}}
\def\poly{{\sf{P}}}

\def\QMA{{\sf{QMA}}}

\def\BQP{{\sf{BQP}}}

\def\Pr{{\text{Pr}}}

\def\hpic #1 #2 {\mbox{$\begin{array}[c]{l}
\epsfig{file=#1,height=#2} \end{array}$}}
\def\vpic #1 #2 {\mbox{$\begin{array}[c]{l}
\epsfig{file=#1,width=#2}\end{array}$}}

\begin{document}
\title{ $\NP$ vs $\QMA_{\log}(2)$}

\author{
Salman Beigi\\ {\it \small Institute for Quantum Information}\\ {\it \small California Institute of
Technology}\\ {\it \small Pasadena, CA}
}

%\author{Salman Beigi\thanks{salman@caltech.edu}  \\
%\normalsize\it{Institute for Quantum Information, California Institute of Technology } }

\date{}

\maketitle{}

\begin{abstract}
Although it is believed unlikely that $\NP$-hard problems admit efficient quantum algorithms, it has been shown
that a quantum verifier can solve $\NP$-complete problems given a ``short" quantum proof; more precisely,
$\NP\subseteq \QMA_{\log}(2)$ where $\QMA_{\log}(2)$ denotes the class of quantum Merlin-Arthur games in which there are
two unentangled provers who send two logarithmic size quantum witnesses to the verifier. The inclusion $\NP\subseteq \QMA_{\log}(2)$
has been proved by Blier and Tapp by stating a quantum Merlin-Arthur protocol for 3-coloring with perfect completeness and gap $\frac{1}{24n^6}$. Moreover, Aaronson {\it et al.}
have shown the above inclusion with a constant gap by
considering $\widetilde{O}(\sqrt{n})$ witnesses of
logarithmic size. However, we still do not know if $\QMA_{\log}(2)$ with a constant
gap contains $\NP$. In this paper, we show that 3-SAT admits a $\QMA_{\log}(2)$ protocol with the gap
$\frac{1}{n^{3+\epsilon}}$ for every constant $\epsilon>0$.

\end{abstract}

\section{Introduction}

$\QMA$ is the class of problems that can be solved by a quantum
polynomial time verifier (Arthur), given a polynomial size
quantum proof by Merlin. The notion of quantum nondeterminism was first
discussed by Knill \cite{knill}, and then studied by Kitaev \cite{kitaev-aqip} and Watrous \cite{watrous}. Later by the profound result
of Kiteav {\it et al.} \cite{kitaev}, who showed that the {\it local
Hamiltonian problem} is $\QMA$-complete, $\QMA$ was turned to an
important complexity class. Although $\QMA$ and the local Hamiltonian
problem are considered as the quantum analogue of $\NP$ and
3-SAT, respectively, there are other types of quantum Merlin-Arthur games without
any classical analogue.

In the classical case, $k$ Merlins, each one of which sends Arthur
his own witness, is the same as one Merlin who sends all the
messages together. However, in the quantum case we may consider the case where the $k$
Merlins are not entangled and then send a separable state to
Arthur. Thus, we cannot argue that one Merlin can send all
the witnesses since he may cheat by sending an entangled state.
So we obtain the non-trivial complexity class $\QMA(k)$ which has been first defined by
Kobayashi {\it et al.} \cite{kobayashi}.

By definition, we have $\QMA=\QMA(1)\subseteq \QMA(2) \subseteq \QMA(3) \subseteq
\cdots$, so a question that arises is that whether we have
equality somewhere or whether all the inclusions are strict. Also,
the gap amplification problem is not an easy one for $\QMA(k)$.
The first idea toward proving gap amplification is to ask each Merlin
to send polynomially many copies of his witness and then repeat
the verification procedure many times. But this idea fails because one of
the Merlins may cheat by entangling his copies. Then after the
first round of the procedure we end up with some entanglement
between different messages, which is not allowed. So there are
two important questions regarding $\QMA(k)$: first, is there some
$k$ such that $\QMA(k+1)=\QMA(k)$, and second, can we amplify the
gap in $\QMA(k)$ protocols? It is interesting that these two
questions are related  \cite{kobayashi, kobayashi2, aaronson}; if we could
amplify the error in $\QMA(k)$ protocols, then $\QMA(2)=\QMA(k)$,
for any $k\geq 2$. Also, it has been proved by Aaronson {\it et al.}
\cite{aaronson} that we can amplify the gap if the {\it Weak
Additivity Conjecture} holds.

Other than changing the number of Merlins, we can consider the case where
the size of the witnesses is less than $\text{poly}(n)$. For instance, in
the classical case $\log(n)$-size witnesses never help the
verifier to solve any problem beyond $\poly$ because he can check
all such witnesses in polynomial time. But this argument fails in
the quantum case and we can define the complexity classes
$\QMA_{\log}(k)$. Although the strong gap amplification
protocol of \cite{marriott} for $\QMA=\QMA(1)$ shows that for
$k=1$ we have $\QMA_{\log}=\BQP$, which is the same situation as
in the classical case, we do not know any non-trivial upper bound
for $\QMA_{\log}(2)$.

Recently, Blier and Tapp \cite{tapp} have shown that $\QMA_{\log}(2)$
with perfect completeness and soundness $1-\frac{1}{24n^6}$
contains the 3-coloring problem, turning this complexity class to an interesting
one which contains both $\BQP$ and $\NP$. The only issue regarding this result is that
the gap should be small ($\frac{1}{24n^6}$). In contrast,
Aaronson {\it et al.} \cite{aaronson} have proved that $\NP$ has a constant gap quantum
Merlin-Arthur protocol in which there are
$\widetilde{O}(\sqrt{n})$ Merlins each one of which sends a
$\log(n)$-qubit state.

In this paper, we show that
3-SAT is in $\QMA_{\log}(2)$ with the gap
$\frac{1}{n^{3+\epsilon}}$ for any constant $\epsilon>0$. Comparing to \cite{tapp}, we improve the gap at the cost of losing perfect completeness.

\subsection{Main idea}

Suppose that Arthur is given a quantum state over two
registers of size $\log(n)$, and wants to recognize whether this
state is entangled or not. We do not know any algorithm to recognize entanglement, but if two unentangled
Merlins give Arthur two witnesses, by comparing them to
his state he can check whether the state is separable or not. It
means that a two-prover Merlin-Arthur protocol can recognize
separable states. On the other hand, Gurvits \cite{gurvits} has
shown that given the classical description of a quantum state over
two registers, it is $\NP$-complete to decide whether the state is
separable or not. Therefore, we have a way of comparing
$\QMA_{\log}(2)$ and $\NP$. This is the main idea behind
our result, but it should be slightly changed in order to
obtain a larger gap.

\section{Definitions and basic properties}

Through this paper we assume the
basic knowledge on theory of quantum computing \cite{chuang} and complexity
theory \cite{sipser, papadim}.

\subsection{$\QMA_{\log}(2)$}

\begin{definition} Let $k$ be an integer, and $a=a(n), b=b(n)$ be
functions such that, $0\leq b< a\leq 1$. Also, let $f(n)$ be a
function of $n$. Then the complexity class $\QMA_{f(n)}(k, a, b)$
consists of languages $L$ for which there exists a quantum
polynomial time verifier $V$ such that for any $x\in \{0,1\}^n$,
\begin{itemize}
\item Completeness: if $x\in L$, then there are $O(f(n))$-qubit
states $\vert \psi_1 \rangle, \dots,\vert \psi_k\rangle$ such
that $\Pr \left[V \text{accepts}\, \vert x\rangle\vert
\psi_1\rangle\dots\vert\psi_k\rangle \right]\geq a$.

\item Soundness: if $x\notin L$, then for any $O(f(n))$-qubit
states $\vert \psi_1 \rangle, \dots,\vert \psi_k\rangle$ we have
\\ $\Pr \left[V \text{accepts}\, \vert x\rangle\vert
\psi_1\rangle\dots\vert\psi_k\rangle \right]\leq b$.
\end{itemize}

\end{definition}

%
%In other words, $\QMA_{f(n)}(k, a,b)$ contains all languages that
%have a quantum Merlin-Arthur protocol with $k$ unentangled
%Merlins.

Here, by convention when the number $k$ or function $f(n)$ are not mentioned we
mean that $k=1$ and $f(n)$ is a polynomial of $n$. Also, we let $\QMA_{f(n)}(k)$ to be
\begin{equation}
\QMA_{f(n)}(k) = \bigcup_{a(n), b(n)} \QMA_{f(n)}(k, a, b),
\end{equation}
where the union is taken over all functions $a(n)$ and $b(n)$ such that $0\leq b(n)< a(n) \leq 1$, and $a(n)-b(n)> n^{-c}$ holds for sufficiently large $n$ and some constant $c$.

Other than the
usual case $f(n)=\text{poly}(n)$, $f(n)=\log(n)$ is also of interest.
Marriott and Watrous \cite{marriott} have considered $f(n)=\log(n)$
for the first time.

\begin{thm} \label{thm:watrous} {\rm \cite{marriott}}
$\QMA_{\log}=\BQP$.
\end{thm}

Proof of this theorem is based on a gap amplification argument
without increasing the size of witness, which is not known for $\QMA(2)$. So we cannot argue that
$\QMA_{\log}(2)$ is the same as $\BQP$. Indeed, it is
a non-trivial complexity class due to the result of Blier and Tapp.

\begin{thm} \label{thm:tapp} {\rm \cite{tapp}}
{\rm 3-coloring}  belongs to $ \QMA_{\log}(2, 1, 1-\frac{1}{24n^6})$.
\end{thm}

\subsection{2-out-of-4-SAT}

To prove the containment $\NP\subseteq \QMA_{\log}(2)$ we should
find a protocol to solve some $\NP$-complete problem in
$\QMA_{\log}(2)$. Although the most well-known such problem is
3-SAT, it is convenient for us to use a variant of this
problem called 2-out-of-4-SAT.

Any instance of 2-out-of-4-SAT consists of some clauses each of which contains exactly four literals, and is satisfied if in each clause exactly two of the literals are true. 2-out-of-4-SAT can also be expressed as
follows.

The clauses of the problem are vectors $\vert a_1\rangle, \vert a_2\rangle, \dots ,\vert
a_m\rangle$ of the form
\begin{equation}\label{eq:4sat}\vert a_k\rangle
=\sum_{j=1}^n c_{kj}\,\vert j\rangle,
\end{equation}
where $c_{kj}= 0$ or $\pm\frac{1}{2}$, and for each $k$ there are
exactly four non-zero $c_{kj}$, $1\leq j\leq n$. We say that the $j$-th variable appears in clause $\vert a_k\rangle$ if $c_{kj}$ is non-zero.
Now the problem
is to decide whether there exists a vector $\vert \psi\rangle$
orthogonal to all $\vert a_k\rangle$'s and of the form
\begin{equation} \label{eq:state}\vert \psi\rangle =\sum_{j=1}^{n}
\pm\frac{1}{\sqrt{n}}\,\vert j\rangle.
\end{equation}

\begin{lem}\label{lem:4-sat} {\rm \cite{aaronson}} There exists a polynomial time Karp
reduction that maps a {\rm 3-SAT} instance $\alpha$ to a {\rm
2-out-of-4-SAT} instance $\beta$ such that
\begin{itemize}

\item If $\alpha$ has $n$ variables and $n'\geq n$ clauses, then
$\beta$ has $O(n'\text{\rm poly}\log (n'))$ variables and $O(n'\text{\rm poly}\log (n'))$
clauses.

\item Every variable of $\beta$ occurs in at most $c$ clauses, for
some constant $c$.

\item The reduction is a PCP, meaning that satisfiable instances
map to satisfiable instances, while unsatisfiable instances map
to instances in which at most a constant fraction of the clauses
can be satisfied at the same time.

\end{itemize}
\end{lem}

\section{Complexity of recognizing entanglement}

Let $H$ be a hermitian matrix of polynomial size (over $\log(n)$ qubits). Then, the
problem of maximizing $\langle \phi \vert H\vert \phi\rangle$
over all states $\vert \phi\rangle$ is an eigenvalue problem and
can be solved efficiently. Now assume that we restrict $\vert \phi\rangle$ to be a separable state.
(Here we assume that $H$ acts over two registers.) Then the above maximization
is an $\NP$-hard problem due to the following observation by
Gurvits  \cite{gurvits}.

Let $H$ be of the form
\begin{equation} \label{eq:matrix} H=\begin{pmatrix}
  0 & B_1 & \cdots & B_{s} \\
  B_1 & 0 & \cdots & 0 \\
  \vdots & \vdots & \ddots & \vdots \\
  B_{s} & 0 & \ldots & 0
\end{pmatrix},
\end{equation}
where $B_j$, $1\leq j\leq s$, is a hermitian matrix. Observe that
$$\langle \psi\vert \langle \phi\vert H\vert \phi\rangle\vert
\psi\rangle = \langle \phi\vert H(\vert\psi\rangle)\vert
\phi\rangle,
$$
where
\begin{equation} \label{eq:matrix} H(\vert \psi\rangle)=\begin{pmatrix}
  0 & \langle \psi\vert B_1\vert \psi\rangle & \cdots & \langle \psi\vert B_{s}\vert \psi\rangle \\
  \langle \psi\vert B_1 \vert \psi\rangle& 0 & \cdots & 0 \\
  \vdots & \vdots & \ddots & \vdots \\
  \langle \psi\vert B_{s} \vert \psi\rangle & 0 & \ldots & 0
\end{pmatrix}.
\end{equation}
This means that the maximum of $\langle \psi\vert \langle \phi\vert
H\vert \phi\rangle\vert \psi\rangle$, for a fixed $\vert
\psi\rangle$, is equal to the maximum eigenvalue of $H(\vert
\psi\rangle)$. $H(\vert \psi\rangle)$ is a rank-two matrix and its eigenvalues can be simply computed. Hence,
\begin{equation} \label{eq:maximum}
\max_{\vert \phi\rangle\vert \psi\rangle}\, \langle \psi\vert
\langle \phi\vert H\vert \phi\rangle\vert \psi\rangle =
\max_{\vert \psi\rangle} \big[ \,\langle \psi\vert B_1\vert
\psi\rangle^2+\cdots +\langle \psi\vert B_s\vert \psi\rangle^2
\,\big]^{1/2}.
\end{equation}
Gurvits \cite{gurvits} has referred to \cite{bental} (which states that estimating the right hand side of Eq. (\ref{eq:maximum}) is $\NP$-hard) and concluded the $\NP$-hardness of computing the left hand side of Eq. (\ref{eq:maximum}).

In this paper, we take the advantage of Eq. (\ref{eq:maximum}) in another direction. Suppose two (unentangled) quantum provers
send the state $\vert \phi\rangle \vert
\psi\rangle$ to a quantum polynomial time verifier. Then the verifier can estimate
$\langle \psi\vert \langle \phi\vert H\vert \phi\rangle\vert
\psi\rangle$ (using the idea of \cite{kitaev, survey}) or equivalently the right hand side of Eq.
(\ref{eq:maximum}). Thus, we conclude that $\QMA_{\log}(2)$ contains $\NP$. Here we slightly change this idea in order to obtain a larger gap in the $\QMA_{\log}(2)$ protocol.

%%%%%%%%%%%%%%%%%%%%%%%%%%%%%%%%%%%%%%%%%%%%%%%%%%%%%%%%

\section{$\NP\subseteq \QMA_{\log}(2)$}

In this section we prove our main result.

\begin{thm}\label{thm:main}
For every constant $\epsilon
>0$, {\rm 3-SAT} is in $\QMA_{\log}(2, a, a-
\frac{1}{n^{3+\epsilon}})$ for some $a$ independent of
$\epsilon$.

\end{thm}

To prove this theorem we give a Merlin-Arthur protocol for the
2-out-of-4-SAT problem. This protocol consists of two parts:
first, given a satisfying assignment we should check whether it is a {\it proper state}, i.e., a state of the form of Eq.
(\ref{eq:state}); second, we should check whether it is orthogonal
to all vectors in the 2-out-of-4-SAT instance. We state each one of these
parts in a separate lemma.

\begin{lem} \label{lem:detect} Let $\epsilon>0$ be a constant. Then
there exists a Merlin-Arthur protocol in which Arthur upon receiving the state
$\vert \phi\rangle \vert \psi\rangle$ can check
whether $\vert \psi\rangle$ is $(5n^{-\epsilon/4})$-close, in
trace distance, to a proper state or not. More precisely, if $\vert
\psi\rangle$ is proper (and $\vert \phi\rangle$ is chosen correctly), then Arthur accepts with probability
\begin{equation} \label{eq:ziad}
\frac{1}{2}+\frac{1}{3n}\left(2-\frac{2}{n}\right)^{1/2},
\end{equation}
and if it is not $(5n^{-\epsilon/4})$-close to a proper state, then he accepts with probability at most
\begin{equation}
\frac{1}{2}+\frac{1}{3n}\left(2-\frac{2}{n}\right)^{1/2} -\frac{1}{20n^{3+\epsilon}}.
\end{equation}

Note that the acceptance probability of this protocol is never more than Eq. (\ref{eq:ziad}).
\end{lem}

Now consider an instance $\alpha$ of 3-SAT. Arthur can reduce
$\alpha$ to an instance $\beta$ of 2-out-of-4-SAT with the
conditions in Lemma \ref{lem:4-sat}, and ask Merlin to send him a
satisfying assignment of $\beta$. Then if $\beta$ is satisfiable, Arthur by
measuring Merlin's state can verify whether it is orthogonal to
$\vert a_k\rangle$'s or not. This idea is elaborated by Aaronson
{\it et al.} \cite{aaronson} to give a protocol for checking whether a given proper
state is a satisfying assignment for $\beta$ or not.

\begin{lem} \label{lem:proper} {\rm \cite{aaronson}} Let us assume
that Merlin is restricted to send a proper state. Then Arthur can
solve {\rm 3-SAT} with perfect completeness and constant
soundness.
\end{lem}

The following corollary is a straightforward consequence of this
lemma.

\begin{corol} \label{corol:1} {\rm \cite{aaronson}} Let us assume that
Merlin is
restricted to send a state that is $\delta$-close, in trace
distance, to a proper state for a constant $\delta>0$. Then
Arthur can solve {\rm 3-SAT} with perfect completeness and
constant soundness.
\end{corol}

Now we prove Theorem \ref{thm:main} assuming Lemma \ref{lem:detect}.\\

\noindent{\bf Proof of Theorem \ref{thm:main}:} Given a 3-SAT
instance $\alpha$ of size $n$ (for a sufficiently large $n$), Arthur reduces it to a 2-out-of-4-SAT instance
$\beta$ over $m$ variables according to Lemma \ref{lem:4-sat}, and asks
Merlins to send him $\vert \phi\rangle \vert \psi\rangle$ where
$\vert \psi\rangle$ is (a proper state and) a satisfying assignment for $\beta$. Then
he applies one of the tests in Lemmas \ref{lem:detect} or
\ref{lem:proper}, each with probability $1/2$.

%If $\vert \psi\rangle$ is not $(5m^{-\epsilon'/4})$-close to a
%proper state (for some $0<\epsilon'<\epsilon$ to be determined), Arthur accepts the
%test in Lemma \ref{lem:detect} with probability at most
%$$\frac{1}{2}+\frac{1}{3m}\left(2-\frac{2}{m}\right)^{1/2}
%-\frac{1}{20m^{3+\epsilon'}}.$$ On the other hand, if $\vert \psi\rangle$ is
%$(5m^{-\epsilon'/4})$-close (and then $2^{-10}$-close for sufficiently large $n$) to a proper state while not a
%satisfying assignment, Arthur rejects the test of Lemma
%\ref{lem:proper} and Corollary \ref{corol:1} with a constant
%probability.

If $\alpha$ is satisfiable, then Arthur accepts with probability
\begin{equation}
a=\frac{1}{2} + \frac{1}{2} \left[\frac{1}{2}+ \frac{1}{3m}\left(2-\frac{2}{m}\right)^{1/2}  \right].
\end{equation}
If it is not satisfiable, then there are two cases. If $\vert \psi\rangle$ is not $(5m^{-\epsilon'/4})$-close to a proper state, then Arthur accepts with probability at most
\begin{equation}
b_1= \frac{1}{2} + \frac{1}{2} \left[  \frac{1}{2}+\frac{1}{3m}\left(2-\frac{2}{m}\right)^{1/2} -\frac{1}{20m^{3+\epsilon'}}\right].
\end{equation}
Also, if $\vert \psi\rangle$ is $(5m^{-\epsilon'/4})$-close (and then $2^{-10}$-close) to a proper state (which is not a satisfying assignment), then he accepts with probability at most
\begin{equation}
b_2= \frac{1}{2} s + \frac{1}{2}\left[\frac{1}{2}+ \frac{1}{3m}\left(2-\frac{2}{m}\right)^{1/2}  \right],
\end{equation}
where the constant $s$ denotes the soundness of the test of Corollary \ref{corol:1} corresponding to $\delta=2^{-10}$. Here we use the fact that the maximum acceptance probability of the protocol of Lemma \ref{lem:detect} is given by Eq. (\ref{eq:ziad}).

Now observe that $b_2<b_1$ for sufficiently large $m$. Therefore, 3-SAT is in $\QMA_{\log}(2, a, b)$, where
\begin{equation}
b= \frac{1}{2}+ \frac{1}{2}\left[\frac{1}{2}+ \frac{1}{3m}\left(2-\frac{2}{m}\right)^{1/2}  \right]- \frac{1}{n^{3+\epsilon}}= a - \frac{1}{n^{3+\epsilon}}.
\end{equation}
Here we replace $\epsilon'$ with $\epsilon$ to consider the
poly-logarithmic blowup in the size of problem by reducing it
from a 3-SAT instance to a 2-out-of-4-SAT instance, and to
eliminate the constants appeared in Lemma \ref{lem:detect}. $\Box$

So the only remaining part is the proof of Lemma \ref{lem:detect}.\\

\subsection{Proof of Lemma \ref{lem:detect}}

Consider a Hilbert space with the orthonormal basis $\{\vert 1\rangle, \dots, \vert n\rangle\}$.
For any $1\leq j < l\leq n$ define the hermitian matrix
$$B_{jl}=\vert j\rangle\langle l\vert+\vert l\rangle\langle j\vert,$$
and let
\begin{equation} \label{eq:h} H=\begin{pmatrix}
  0 &  B_{1,2} & \cdots &  B_{(n-1)n}  \\
   B_{1,2} & 0 & \cdots &  0 \\
  \vdots & \vdots & \ddots & \vdots \\
   B_{(n-1)n}  & 0 & \cdots   & 0
\end{pmatrix},
\end{equation}
where all $B_{jl}$, $1\leq j < l\leq n$, appear as a submatrix of $H$.

We show that the maximum of $\langle \psi\vert \langle \phi\vert H\vert \phi\rangle \vert \psi\rangle$ over all states $\vert \psi\rangle$ and $\vert \phi\rangle$ occurs when $\vert \psi\rangle$ is a proper state. In this case, given the state $\vert \phi\rangle \vert \psi\rangle$ one can estimate $\langle \psi\vert \langle \phi\vert H\vert \phi\rangle \vert \psi\rangle$ in order to check whether $\vert \psi\rangle$ is a proper state or not. However, $H$ is not a measurement operator and it is not clear how we can estimate $\langle \psi\vert \langle \phi\vert H\vert \phi\rangle \vert \psi\rangle$. So we need some modifications.

It is easy to see that $\lambda\neq 0$
is an eigenvalue of $H$ iff $\lambda^2$ is an eigenvalue of
$\sum_{j,l} B_{jl}^2$. Then, $\| H\|_{\infty}$, the infinite-norm\footnote{$\|X\|_{\infty}$ denotes the maximum eigenvalue
of $\vert X\vert=\sqrt{XX^{\dagger}}$.} of matrix $H$, satisfies
$$\|H \|_{\infty}^2 = \|\sum_{j,l} B_{jl}^2 \|_{\infty}\leq \sum_{j,l} \|B_{jl}\|_{\infty}^2 = {n \choose{2}}  \leq n^2.$$
Therefore, $\frac{1}{2}I+\frac{1}{3n} H$ is a positive
semi-definite matrix (and in fact an $O(\log(n))$-local
Hamiltonian) with norm  $\| \frac{1}{2}I+\frac{1}{3n} H \|_{\infty}< 1$.
Thus, by the techniques presented in \cite{kitaev,survey}, having the state $\vert
\phi\rangle\vert \psi\rangle$ Arthur can throw a coin with
probability of head being
\begin{equation}
\langle \psi\vert \langle \phi\vert
\big(\frac{1}{2}I+\frac{1}{3n} H\big) \vert \phi\rangle \vert \psi\rangle,
\end{equation}
and accept if it is head. Hence, by Eq. (\ref{eq:maximum}), if $\vert \phi\rangle$ is the right state, the
probability of acceptance is equal to
\begin{equation} \label{eq:1/n}\frac{1}{2}+\frac{1}{3n} \max_{\vert
\psi\rangle} \left[\, \sum_{j,l} \langle \psi\vert B_{jl}\vert
\psi\rangle^2\, \right]^{1/2}.
\end{equation}
Now we need the following lemma.

\begin{lem} \label{lem:close} $\sum_{j,l} \langle \psi\vert B_{jl}\vert
\psi\rangle^2 \leq  2-\frac{2}{n}$, and equality holds iff $\vert
\psi\rangle $ is a proper state. Also, for sufficiently large $n$ if
\begin{equation} \label{eq:ineq}\sum_{j,l} \langle \psi\vert B_{jl}\vert \psi\rangle^2 \geq
2-\frac{2}{n}- \frac{1}{n^{2+\epsilon}}, \end{equation} then
$\vert \psi \rangle$ is $(5n^{-\epsilon/4})$-close to a proper
state in trace distance.

\end{lem}

Using this lemma, if $\vert \psi \rangle$ is a proper state, the probability of acceptance is equal to
\begin{equation}\frac{1}{2}+ \frac{1}{3n}\left(2-\frac{2}{n}\right)^{1/2},
\end{equation} and if it is
greater than
\begin{equation}\frac{1}{2}+ \frac{1}{3n}\left(2-\frac{2}{n}\right)^{1/2}- \frac{1}{20n^{3+\epsilon}},
\end{equation}
then $\vert \psi \rangle$ is
$(5n^{-\epsilon/4})$-close to a proper state. $\Box$

So it remains to prove Lemma \ref{lem:close}.\\

\noindent{\bf Proof of Lemma \ref{lem:close}:} Let
\begin{equation}\vert
\psi\rangle =\sum_{j=1}^{n} \,x_j\, \vert j\rangle
\end{equation} be a normalized state. Then

\begin{eqnarray} \sum_{j,l} \, \langle \psi\vert B_{jl}\vert
\psi\rangle^2 & =  &\sum_{j<l}\
(\overline{x}_jx_l+x_j\overline{x}_l)^2 \nonumber\\
& = & \sum_{j<l}\
\big(\overline{x}_j^2x_l^2+x_j^2\overline{x}_l^2+2\vert
x_j\vert^2\vert x_l\vert^2 \big)\nonumber\\
& = & \big(\sum_j\ \overline{x}_j^2\big)\big(\sum_j\ x_j^2\big)
-\sum_j \vert
x_j\vert^4+2\sum_{j<l} \vert x_j\vert^2\vert x_l\vert^2 \nonumber\\
& =& \vert\sum_j\ x_j^2\vert^2+ \big(\sum_j \vert
x_j\vert^2\big)^2-2\sum_j \vert
x_j\vert^4,\label{eq:eqn1}
\end{eqnarray}
where $\overline{x}$ denotes the complex conjugate of the number $x$.

Using equation $\sum_{j=1}^{n}\vert x_j\vert^2=1$ we obtain the inequalities
\begin{equation} \label{eq:in1}
\sum_j \vert x_j\vert^4 \geq \frac{1}{n}
\end{equation}
and
\begin{equation}\label{eq:in2}
\vert\sum_j\ x_j^2\vert^2 \leq 1.
\end{equation}
Hence, combining with Eq. (\ref{eq:eqn1}) we find that $\sum_{j,l} \, \langle \psi\vert B_{jl}\vert
\psi\rangle^2 \leq 2-\frac{2}{n}$, and equality holds iff both Eqs. (\ref{eq:in1}) and (\ref{eq:in2}) are equalities, i.e., for any $1\leq j\leq n$
\begin{equation}
x_j^2=\frac{1}{n}\,e^{i\theta},
\end{equation}
for a constant
$\theta$, or equivalently iff $\vert \psi\rangle$ is a proper
state.

Now assume that Eq. (\ref{eq:ineq}) holds; we show that $\vert \psi\rangle$ is close to a proper state. By Eq. (\ref{eq:eqn1}) we have
\begin{equation}
\sum_{j,l} \langle \psi\vert B_{jl}\vert \psi\rangle^2= \vert\sum_j\ x_j^2\vert^2+ \big(\sum_j \vert
x_j\vert^2\big)^2-2\sum_j \vert
x_j\vert^4 \geq
2-\frac{2}{n}- \frac{1}{n^{2+\epsilon}}.
\end{equation}
So comparing to Eqs. (\ref{eq:in1}) and
(\ref{eq:in2}) we find that
\begin{equation} \label{eq:app1}
\sum_j \vert x_j\vert^4 \leq \frac{1}{n}+\frac{1}{n^{2+\epsilon}},
\end{equation}
and
\begin{equation}\label{eq:app2}
\vert\sum_j\ x_j^2\vert^2 \geq 1 -\frac{1}{n^{2+\epsilon}}.
\end{equation}

Observe that
\begin{equation} \sum_j \left(\vert x_j\vert^2 -\frac{1}{n}\right)^2= \sum_j \big( \vert x_j\vert^4 + \frac{1}{n^2}
 - \frac{2}{n}\vert x_j\vert^2 \big) = \sum_j \vert x_j\vert^4 -\frac{1}{n}.
\end{equation}
Therefore, by Eq. (\ref{eq:app1}) for every $j$
\begin{equation}\label{eq:norm1}
\vert \,\vert x_j\vert^2 -\frac{1}{n}\, \vert\leq
\frac{1}{n^{1+\delta}},
\end{equation}
where $\delta=\epsilon/2$, and then
\begin{equation} \label{eq:norm2}
\vert \, \vert x_j\vert -\frac{1}{\sqrt{n}}\, \vert \leq
\frac{\sqrt{n}}{n^{1+\delta}}.
\end{equation}
Also, using Eq. (\ref{eq:app2}) we have
\begin{equation}
\big( \, \sum_j \vert x_j\vert^2\,\big)^2 - \vert \, \sum_j x_j^2\,
\vert^2 \leq \frac{1}{n^{2+\epsilon}},
\end{equation}
and since \begin{equation}(\sum_j|x_j|^2)^2 - |\sum_j x_j^2|^2
 = 2\sum_{ j<l}\, (|x_j x_l|^2 - \text{Re}\, x_j^2 \bar{x}_l^2),\end{equation}
and $|x_jx_l|^2 -\text{Re}\, x_j^2\bar{x}_l^2$ is always non-negative
we obtain
\begin{equation}\label{eq:angle}
\vert x_jx_l\vert^2 -\text{Re}\, x_j^2\overline{x}_l^2 \leq
\frac{1}{n^{2+\epsilon}},
\end{equation}
for every $j$ and $l$.

Now let $x_j= s_jr_je^{i\theta_j}$, where $s_j\in \{+1, -1\}$,
$r_j$ is a non-negative real number, and $-\frac{\pi}{2} <
\theta_j\leq \frac{\pi}{2}$. Then Eq. (\ref{eq:norm2}) is equivalent to
\begin{equation}\label{eq:r1}
\vert r_j - \frac{1}{\sqrt{n}}\,\vert \leq
\frac{\sqrt{n}}{n^{1+\delta}}.
\end{equation}
Also, by Eqs. (\ref{eq:norm1}) and (\ref{eq:angle})
\begin{equation} \label{eq:theta1}
1- \text{Re}\, e^{2i(\theta_j-\theta_l)} \leq
\frac{1}{n^{2+\epsilon}}\big( \frac{1}{n} -\frac{1}{n^{1+\delta}}
\big)^{-2} = (n^\delta - 1)^{-2} \leq \frac{2}{n^{\epsilon}},
\end{equation}
for sufficiently large $n$. Without loss of generality, we
assume that $\theta_1=0$; thus for every $j$ we have
\begin{equation} \label{eq:theta2}
1- \text{Re}\, e^{2i\theta_j}  \leq \frac{2}{n^{\epsilon}},
\end{equation}
and since $-\frac{\pi}{2}<\theta_j \leq \frac{\pi}{2},$
\begin{equation} \label{eq:theta3}
1- \text{Re}\, e^{i\theta_j}  \leq \frac{2}{n^{\epsilon}}.
\end{equation}
Now using $(\text{Re}\, e^{i\theta_j})^2 + (\text{Im}\, e^{i\theta_j})^2=1$, it
is easy to see that
\begin{equation} \label{eq:theta4}
\vert 1 - e^{i\theta_j}\, \vert \, \leq \frac{2}{n^\delta}.
\end{equation}
Therefore, by Eqs. (\ref{eq:r1}) and (\ref{eq:theta4})
\begin{eqnarray}
\vert\,  r_je^{i\theta_j} - \frac{1}{\sqrt{n}}  \,\vert & \leq &
\vert r_j -\frac{1}{\sqrt{n}}\vert + \vert r_i (1- e^{i\theta_j}
)\vert \nonumber\\
& \leq  & \frac{\sqrt{n}}{n^{1+\delta}}  + \left(
\frac{1}{\sqrt{n}}+\frac{\sqrt{n}}{n^{1+\delta}} \right)
\frac{2}{n^\delta},
\end{eqnarray}
and then
\begin{equation} \label{eq:rtheta}
\vert\,  r_je^{i\theta_j} - \frac{1}{\sqrt{n}}  \,\vert \leq
\frac{10\sqrt{n}}{n^{1+\delta}}.
\end{equation}

Now define the proper state
\begin{equation}\vert \psi'\rangle = \sum_j \frac{s_j}{\sqrt{n}} \, \vert j\rangle.
\end{equation}
We have
\begin{eqnarray}
\vert\, \langle \psi'\vert \psi\rangle\, \vert  & = & \vert \,
\sum_j \frac{1}{\sqrt{n}}\, s_j^2r_je^{i\theta_j} \,\vert \nonumber\\
& = & \frac{1}{\sqrt{n}}\, \vert \,
\sum_j \, r_je^{i\theta_j} \,\vert \nonumber\\
& \geq & \frac{1}{\sqrt{n}}\, \left( \sqrt{n} - \vert \, \sum_j \,
\big( r_je^{i\theta_j} -\frac{1}{\sqrt{n}}\big) \,\vert \right)\nonumber\\
& \geq & 1 - \frac{1}{\sqrt{n}}\sum_j |r_j e^{i \theta_j} - \frac{1}{\sqrt{n}}|.
\end{eqnarray}
Using Eq. (\ref{eq:rtheta}) we obtain
\begin{equation} \label{eq:last}
\vert\, \langle \psi'\vert \psi\rangle\, \vert  \geq 1-
\sqrt{n}\,\frac{10\sqrt{n}}{n^{1+\delta}} = 1-
\frac{10}{n^\delta}.
\end{equation}
Therefore,
\begin{equation}
\|\, \vert \psi\rangle\langle \psi\vert -\vert \psi'\rangle\langle \psi'\vert \,\|_{\text{tr}} = \left( 1- \vert
\langle \psi'\vert \psi\rangle \vert^2 \right)^{1/2} \leq \left(
\frac{20}{n^{\delta}} \right)^{1/2} < 5n^{-\delta/2}.
\end{equation}
We are done. $\Box$

\section{Conclusion}

Although the gap in our $\QMA_{\log}(2)$ protocol for 3-SAT is
larger than the gap in the proof of Blier and Tapp (
$\frac{1}{n^{3+\epsilon}}$ versus $\frac{1}{24n^6}$), their
protocol is one-sided error. So one direction to improve this
result is to turn it into a protocol with perfect completeness.

Another open question is that whether the optimal gap depends on $n$, or
whether there exists a constant gap $\QMA_{\log}(2)$ protocol for
$\NP$. This question is related to the problem of whether
recognizing states that are $\delta$-close to a separable state,
for some {\it constant} $\delta>0$, is $\NP$-hard or not.\\

\noindent{\bf Acknowledgements.} The author is thankful to Peter W.
Shor for helpful discussions. He is also grateful to unknown referees for reporting some errors in the earlier version of this paper.


\small
\begin{thebibliography}{10}
\vspace*{.04in}

\bibitem{knill} E. Knill. Quantum randomness and nondeterminism. Technical Report LAUR-96-2186, Los
Alamos National Laboratory, 1996. quantph/9610012.

\bibitem{kitaev-aqip} A. Kitaev. Quantum NP. Talk at AQIP'99: Second Workshop on Algorithms in Quantum
Information Processing, 1999.

\bibitem{aaronson} S. Aaronson, S. Beigi, A. Drucker, B. Fefferman and P. Shor,
{\it The Power of Unentanglement}, Theory of
Computing, 5 (2009) pp. 1-42


\bibitem{survey} Dorit Aharonov and Tomer Naveh, {\it Quantum NP - A
Survey}, quant-ph/0210077

\bibitem{bental} A. Ben-Tal and A. Nemirovski, {\it Robust convex optimization},
Mathematics of Operational Research, Vol. 23, 4 (1998), 769-805.


\bibitem{tapp} Hugue Blier and Alain Tapp, {\it All languages in NP have very short quantum proofs},
Proceedings of the ICQNM, 2009, 34-37



\bibitem{gurvits} Leonid Gurvits, {\it Classical complexity and quantum
entanglement}, Journal of Computer and System Sciences, 69 (2004) 448–-484



\bibitem{kitaev} A. Kitaev, A. Shen, and M. N. Vyalyi, {\it Classical
and Quantum Computation}, American Mathematical Society, 2002


\bibitem{kobayashi} H. Kobayashi, K. Matsumoto and T. Yamakami, {\it Quantum Certificate Verification: Single versus
Multiple Quantum Certificates}, quant-ph/0110006

\bibitem{kobayashi2} H. Kobayashi, K. Matsumoto and T.
Yamakami, {\it Quantum Merlin-Arthur Proof Systems: Are Multiple
Merlins More Helpful to Arthur?}, Lecture Notes in Computer Science, 2003, vol. 2906, pp. 189-198


\bibitem{marriott} Chris Marriott, John Watrous, {\it Quantum Arthur-Merlin
Games}, Computational Complexity, 14(2): 122-152, 2005

\bibitem{chuang} M. A. Nielsen and I. L. Chuang, {\it Quantum Computation and Quantum
Information}, Cambridge University Press, Cambridge, 2000

\bibitem{papadim} C. H, Papadimitriou, {\it Computational Complexity},
Addison-Wesley Publishing Company, Inc., 1994.

\bibitem{sipser} Michael Sipser, {\it Introduction to the Theory of
Computation}, PWS Publishing Company, 2005

\bibitem{watrous} J. Watrous, {\it Succinct quantum proofs for properties
of finite groups}, Proceedings of IEEE FOCS'2000, pp. 537-546,
2000



\end{thebibliography}
\end{document}